\documentclass[a4paper, 10pt, conference]{ieeeconf} 
\IEEEoverridecommandlockouts
\usepackage[normalem]{ulem}
\usepackage{graphicx}
\usepackage{epstopdf}
\usepackage[caption=false]{subfig}

\usepackage{amsmath,amssymb,amsfonts,amsthm}

\usepackage[ruled,vlined]{algorithm2e}

\DeclareMathOperator*{\argmin}{arg\,min}

\DeclareMathOperator*{\diag}{diag}

\newtheorem{theorem}{Theorem}
\newtheorem{corollary}{Corollary}
\newtheorem{proposition}{Proposition}
\newtheorem{remark}{Remark}

\usepackage[dvipsnames]{xcolor}

\definecolor{light}{rgb}{0.5, 0.5, 0.5}

\begin{document}
\title{A Contraction-constrained Model Predictive Control for Nonlinear Processes using Disturbance Forecasts \thanks{This work was partially supported by Australian Research Council Discovery Project DP210101978.}}

\author{Ryan~McCloy,~Lai Wei,~Jie~Bao
\thanks{R. McCloy, L. Wei, and J. Bao are with the School of Chemical Engineering, The University of New South Wales, Sydney, NSW 2052, Australia. E-mail: j.bao@unsw.edu.au (corresponding author).}}
\maketitle
\begin{abstract}
Model predictive control (MPC) has become the most widely used advanced control method in process industry. In many cases, forecasts of the disturbances are available, e.g., predicted renewable power generation based on weather forecast. While the predictions of disturbances may not be accurate, utilizing the information can significantly improve the control performance in response to the disturbances.  By exploiting process and disturbance models, future system behaviour can be predicted and used to optimise control actions via minimisation of an economical cost function which incorporates these predictions. However, stability guarantee of the resulting closed-loop system is often difficult in this approach when the processes are nonlinear. Proposed in the following article is a contraction-constrained predictive controller which optimises process economy whilst ensuring stabilisation to operating targets subject to disturbance measurements and forecasts.
\end{abstract}

\section{Introduction}
The incorporation of external variable forecasts into advanced control strategies for industrial processes has been well established as economically advantageous. In addition, predictive control strategies are often more efficient, outperforming conventional, non-predictive control strategies, due to their ability to consider future disturbances \cite{RaM09}.  

In process industries, feedforward control is used routinely to compensate for measured disturbances when they are detected \cite{altmann2005practical}, but the standard implementation of feedforward control does not include a prediction of the future evolution of the disturbance. Model Predictive Control (MPC) \cite{Qin2003,RaM09} exploits process models to predict system evolution and optimise control actions via minimisation of a cost function which incorporates these predictions \cite{rawlings2017model}. In many cases, forecasts of the disturbances are available, e.g., predicted renewable power generation based on weather forecast, hourly or daily wastewater discharge patterns based on historical data, etc.  As such, the use of MPC for control of processes whilst considering disturbance forecasts to improve controller performance has become widely popular, for example in smart energy systems (building climate control \cite{oldewurtel2012use} and HVAC systems \cite{mahdavi2017model}, energy storage \cite{teleke2010optimal} and power scheduling \cite{qi2011supervisory} in networks incorporating renewable sources), water networks (irrigation \cite{delgoda2016irrigation}, drinking water management \cite{wang2015robust}, wastewater treatment \cite{stentoft2020integrated}), and even in supply chain management \cite{wang2008model} and automotive industries (active suspension \cite{theunissen2021preview}, electric vehicle energy management \cite{sun2015velocity}).


Typically, the cost function is chosen in MPC such that the optimal cost forms a Lyapunov function for the closed-loop system, resulting in a certificate for stability. This however is not practical when considering cost functions as representations of process economy, whereby operating targets can vary away from those associated with minimal cost. Moreover, in practice, this requirement is largely ignored when considering disturbance forecasts, and whilst defensible for stable systems with slow dynamics, in general, a stabilising condition (both in terms of process safety and operating target tracking) would be considered essential (see, e.g., \cite{ellis2017economic}). Since economic cost functions are generally not positive definite, many existing stability proofs in traditional MPC (e.g.,~\cite{RaM09}) cannot be applied. This has led to the Lyapunov-based MPC designs \cite{heidarinjad2012EMPC}, whereby the economic optimisation problem is solved subject to an additional stability constraint, and closed-loop stability is explicitly ensured by offline Lyapunov based control design. However, as the stability condition is only valid for a specific equilibrium, the control algorithm for dynamic operating targets (which can result due to operating conditions subject to disturbances) requires offline redesign (e.g., with a new Lyapunov function) when the target is updated \cite{ellis2017economic}. Due to this inflexibility with respect to reference changes, the Lyapunov-based MPC approach is impractical for use with variable disturbance forecasts, which require a condition that is reference-independent, e.g., those based on incremental stability \cite{angeli2002lyapunov,santoso2012}. As a consequence, an effective MPC with disturbance forecasts requires an incremental stability constraint or equivalent (analogously to the Lyapunov-based MPC approach in \cite{ellis2017economic}). Introduced by \cite{lohmiller1998contraction}, contraction theory facilitates stability analysis and control of nonlinear systems with respect to arbitrary, time-varying (feasible) references without redesigning the control algorithm \cite{manchester2017control,lopez2019contraction}. One useful feature of contraction theory is that it can be used to analyse the incremental stability of nonlinear systems and simultaneously synthesise a controller that ensures offset free tracking of feasible target trajectories using control contraction metrics (see, e.g., \cite{manchester2017control}). This has motivated developments for contraction-based MPCs \cite{mccloy2021differential,xiao2021robust}, offering significant flexibility over Lyapunov-based alternatives. 

We propose herein a novel contraction constrained MPC approach for discrete-time nonlinear processes subject to predicted disturbances (as exogenous inputs). As disturbance predictions are never accurate, it is also important to attenuate the effects of disturbance prediction errors. Through the contraction theory framework, stability conditions are derived, to ensure convergence of the resulting closed-loop system, i.e., tracking of feasible time-varying references under bounded disturbance measurement error, where the $\mathcal{L}_2$ gain from the disturbance prediction error to the state error is attenuated.  These conditions are then reformed to characterise a set of stabilising controllers with the contracting property and imposed as stability constraints on an MPC. The MPC then solves an optimisation problem to find the most economical controller amongst those stabilising controllers, whilst utilising both disturbance forecasts and measurements. The resulting MPC is capable of ensuring safe economical objective tracking for nonlinear processes subject to measured and forecasted disturbances. 

The remainder of this article is structured as follows, Section \ref{sec:pre} presents the prerequisite contraction theory tools, and Section \ref{sec:pro} formulates the constrained MPC problem. Section \ref{sec:ctr} develops contraction analysis and controller structures for systems subject to disturbances. Section \ref{sec:exa} demonstrates the overall method via numerical simulation and Section \ref{sec:conclusion} concludes this article. 


\section{Preliminaries} \label{sec:pre}
Firstly, we consider the discrete-time nonlinear control affine system without disturbance (extended in later sections)
\begin{equation}\label{equ:pre cer sys}
    x_{k+1} = f(x_k) + g(x_k)u_k,
\end{equation}
where state and control are $x_k \in \mathcal{X} \subseteq \mathbb{R}^n$ and $u_k \in \mathcal{U} \subseteq \mathbb{R}^m$. The corresponding differential system of \eqref{equ:pre cer sys} is

\begin{figure}
    \begin{center}
        \includegraphics[width=\linewidth]{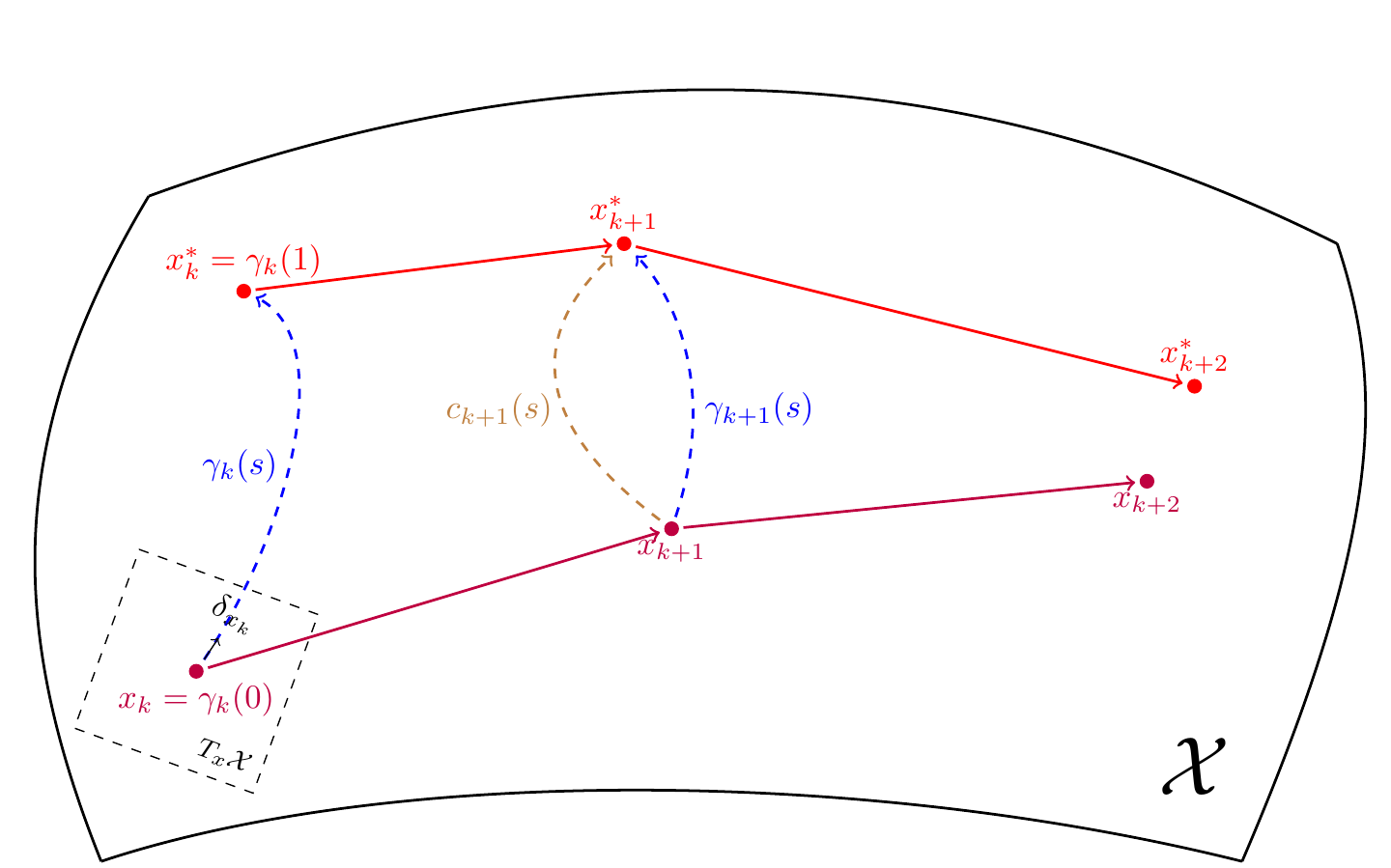}
        \caption{State and Reference Trajectories with $s$-parameterised Paths.}
        \label{fig:rie_geo}
    \end{center}
\end{figure}
\begin{equation}\label{equ:pre cer dif sys}
    \delta_{x_{k+1}} = A\delta_{x_k} + B\delta_{u_k},
\end{equation}
where Jacobian matrices of $f$ and $g$ in \eqref{equ:pre cer sys} are defined as $A:=\frac{\partial (f(x_k) + g(x_k)u_k)}{\partial x_k}$ and $B:=\frac{\partial (f(x_k) + g(x_k)u_k)}{\partial u_k}$  respectively, $\delta_{u_k} := \frac{\partial u_k}{\partial s}$ and $\delta_{x_k}:=\frac{\partial x_k}{\partial s}$ are vectors in the tangent space $T_x\mathcal{U}$ at $u_k$ and tangent space $T_x\mathcal{X}$ at $x_k$ respectively, where $s$ parameterises a path, $c(s): [0,1] \rightarrow \mathcal{X}$ between two points such that $c(0) = x, c(1) = x^*  \in \mathcal{X}$ (see Fig. \ref{fig:rie_geo}). If we consider a state-feedback control law for the differential dynamics \eqref{equ:pre cer dif sys}, i.e.,
\begin{equation}\label{equ:pre dif fed}
    \delta_{u_k} = K(x_k) \delta_{x_k},
\end{equation}
where $K$ is a state dependent feedback gain. Then, Theorem \ref{thm:pre ctr con} \cite{wei2021control} describes the contraction condition \eqref{equ:pre cer sys} as follows.
\begin{theorem}\label{thm:pre ctr con}
    For a discrete-time nonlinear system \eqref{equ:pre cer sys}, with differential dynamics \eqref{equ:pre cer dif sys} and differential state-feedback controller \eqref{equ:pre dif fed}, provided a uniformly metric, $\underline{m}I \leq M(x_k) \leq \overline{m}I$, exists satisfying, 
    \begin{equation}\label{equ:pre ctr con} 
        (A+BK)^\top M_+(A+BK) - (1-\beta)M < 0,
    \end{equation}
    where $M_+=M(x_{k+1})$, the closed-loop system is contracting for some constant $0 < \beta \leq 1$. Furthermore, the closed-loop system is exponentially incrementally stable,  i.e.,
    \begin{equation}
        \label{equ:pre exp sta}
        |x_k-x^*_k| \leq  R e^{-\lambda k} |x_0 - x^*_0|,
    \end{equation}    
    for some constant $R$, convergence rate $\lambda$ and any feasible reference trajectory $(x^*, u^*)$, satisfying \eqref{equ:pre cer sys}, where $x_0$ is the initial state and $x_k$ is the state at time-step $k$.

\end{theorem}
A state space region is called a contraction region if condition \eqref{equ:pre ctr con} holds for all points in that region. In Theorem \ref{thm:pre ctr con}, the metric, $M$, is used in describing such regions using the Riemannian geometry, which we briefly present here. We define the Riemannian distance, $d(x,x^*)$, as (see, e.g., \cite{do1992riemannian})
\begin{equation}\label{equ:Riemannian distance and energy}
    \begin{aligned}
    d(x,x^*) = d(c) :=\int_0^1 \sqrt{\delta^\top_{c(s)}M(c(s))\delta_{c(s)}}ds,
    \end{aligned}
\end{equation}
where $\delta_{c(s)} := \frac{\partial c(s)}{\partial s}$.
The shortest path in Riemannian space, or \textit{geodesic}, between $x$ and $x^*$ is defined as 
\begin{equation}\label{equ:geodesic}
    \gamma(s) :=\argmin_{c(s)} {d(x,x^*)}.
\end{equation}

Leveraging Riemannian tools, a feasible feedback tracking controller for  \eqref{equ:pre cer sys}, can be obtained by integrating the differential feedback law \eqref{equ:pre dif fed} along the geodesic, $\gamma(s)$ \eqref{equ:geodesic}, as
\begin{equation}\label{equ:pre ctl int}
    u_k = u^*_k + \int_0^1K(\gamma(s))\frac{\partial \gamma(s)}{\partial s}\,ds.
\end{equation}

In summary, a suitably designed contraction-based controller ensures that the length of the minimum path (i.e., geodesic) between any two trajectories (e.g., the plant state, $x$, and desired state, $x^*$, trajectories), with respect to the metric $M$, shrinks with time, i.e., provided that the contraction condition \eqref{equ:pre ctr con} holds for the discrete-time nonlinear system \eqref{equ:pre cer sys}, 
we can employ a stabilizing feedback controller \eqref{equ:pre ctl int} to ensure convergence to feasible operating targets.

\section{Problem Formulation} \label{sec:pro}
\subsection{Nonlinear System with Exogenous Input Description}
Consider the discrete-time nonlinear process model with exogenous input (e.g., disturbances), $\nu \in \mathcal{V}$, described as
\begin{equation}\label{eq:sys_dist_nom}
    x_{k+1} = f(x_k) + B_u u_k + B_\nu \nu_k
\end{equation}
with differential dynamics
\begin{equation}\label{eq:dsys_dist_nom}
    \delta_{x_{k+1}} = A\delta_{x_k} + B_u\delta_{u_k} + B_\nu\delta_{\nu_k},
\end{equation}
where $A := \frac{\partial f}{\partial x}$ and $\delta_{\nu_k}:=\frac{\partial \nu_k}{\partial s}$ is a vector in the tangent space $T_x\mathcal{V}$ at $\nu_k$. Here, $s$ parameterises all potential disturbance trajectories analogously to $x(s)$ in Fig. \ref{fig:rie_geo}, i.e., using $\nu(s)$, the actual disturbance is denoted as $\nu(0) = \nu_k$ (at $s=0$) and the modeled (or measured) disturbance is denoted as $\nu(1) = \nu^*_k$ (at $s=1$). 
An $H$-step disturbance forecast is assumed available, denoted as $\boldsymbol{\hat{\nu}} := (\hat{\nu}_k,\hat{\nu}_{k+1},\cdots,\hat{\nu}_{k+H-1})$, and is also assumed updated at each sampling instant, $k$, to include the disturbance measurement, $\nu^*_k$, i.e., $\boldsymbol{\hat{\nu}} = (\nu^*_k,\hat{\nu}_{k+1},\cdots,\hat{\nu}_{k+H-1})$ .

\subsection{MPC using Disturbance Forecasts}
To utilise the disturbance forecast information  effectively, whilst tracking time-varying operating targets, i.e. $x_k\rightarrow x_k^*$, the incorporation of a prediction based control strategy, such as MPC, is naturally befitting. The controller, $u_k$, can then be computed to additionally optimise system economy  via the minimisation of an arbitrary cost function, $\ell(k+i)$, i.e., an MPC solves the following optimisation problem,
\begin{equation}
\begin{aligned}\label{eq:EMPCini}
\quad \min_{\hat{u}} & \sum_{i=0}^{N} \ell (\hat{x}_i,\hat{u}_i), \\
\text{s.t.} \quad & \hat{x}_0 = x_k, \quad \hat{\nu}_0 = \nu_k, \\
\quad & \hat{x}_{i+1} = f(\hat{x}_i) + B_u\hat{u}_i + B_\nu\hat{\nu}_i, \\
& \hat{u}_i \in \mathcal{U}, \quad \hat{x}_i \in \mathcal{X}, \quad r(\hat{x}_i,x^*_i,\hat{\nu}_i,\hat{u}_i,\beta) \leq 0
\end{aligned}
\end{equation}
where $N$ is the prediction horizon, $\hat{x}_i$, $\hat{\nu}_i$ and $\hat{u}_i$ are the respective $i$-th step state, disturbance and control predictions, $x_k$ is the current state measurement, $\mathcal{U},\mathcal{X}$ are the sets of permissible inputs and states respectively, and $r(\hat{x}_i,x^*_i,\hat{\nu}_i,\hat{u}_i,\beta)$ defines a stability constraint. Assuming feasibility of the optimisation problem~\eqref{eq:EMPCini}, i.e., the optimal input trajectory, $\mathbf{\hat{u}}^{opt} = (\hat{u}^{opt}_{0}, \hat{u}^{opt}_{1}, \cdots, \hat{u}^{opt}_{N-1}) \in \mathcal{U}^{N}$, can be computed satisfying \eqref{eq:EMPCini}. The MPC is then implemented in receding horizon fashion, by applying the first control action $\hat{u}^{opt}_{0}$ to system \eqref{eq:sys_dist_nom} until the next step, $k+1$, i.e.,
\begin{equation}
\label{eq:EMPCu}
u(k) = \hat{u}^{opt}_{0} : [k,k+1).
\end{equation}

Since arbitrary cost functions are considered, a stability condition is required \cite{ellis2017economic}. As, the reference trajectory is considered time-varying, stability guarantees for an optimal controller require a condition that is also reference-independent (incremental stability). In the case of MPC, a cost function can be utilised to represent the process economy. Optimal solutions can therefore be associated with state values which deviate from target operating points. As a consequence, an MPC requires a contraction based constraint or equivalent, as will be demonstrated in following sections.

\subsection{Objective and Approach}
To achieve the objective of state trajectory tracking, i.e., $x_k \to x^*_k$, whilst optimising system economy, for nonlinear processes subject to known exogenous inputs (with predictable trajectories), we require an MPC with conditions that can ensure convergency along the full prediction horizon. To this effect, we propose to:
\begin{enumerate}
    \item[i.] Determine tractable stabilising controller conditions using contraction metrics
    \item[ii.] Impose stability conditions as constraints on an MPC
\end{enumerate}
In this way, tractable offline stability conditions can be constructed and imposed online on an MPC, which has the capacity for utilising predicted input and state reference trajectories whilst optimising an economic cost.  

\section{Control Contraction Analysis of Processes with Measured Disturbances}\label{sec:ctr}
In the following section, the contraction theory framework will be leveraged to derive stability conditions to ensure objective convergence of the resulting closed-loop system under bounded disturbance measurement error. These conditions are transformed into tractable synthesis conditions and ultimately used to characterise the property of stabilising controllers forming the necessary MPC stability constraints. 
\subsection{Analysis under Correct Disturbance Prediction}
\begin{proposition}\label{prop:ccon_no synth}
Provided the contraction condition
\begin{equation}\label{eq:dist ccon} 
\begin{split}
    \delta_{x_k}^\top A_{c}^\top M_+ A_{c}\delta_{x_k} - (1-\beta)\delta_{x_k}^\top M \delta_{x_k}
    \\+ \delta_{x_{k}}^\top A_{c}^\top M_+B_{\nu}\delta_{\nu_k} 
    + \delta^\top_{\nu_k} B^\top_{\nu}M_+A_{c}\delta_{x_{k}}
    \\+ \delta^\top_{\nu_k} B^\top_{\nu}M_+B_{\nu}\delta_{\nu_k} 
      < 0
\end{split}
\end{equation}
where $A_c = A(x_k)+B_u K(x_k)$, $M_+=M(x_{k+1})$, can be satisfied for the pair $(M,K)$ $\forall x,\delta_x \in \mathcal{X}$, $\nu,\delta_\nu \in \mathcal{V}$, \eqref{eq:sys_dist_nom} is exponentially stabilisable to every feasible trajectory $x^*_k$ with respect to the metric $M$.
\end{proposition}
\begin{proof}
Suppose there exists a smooth feedback controller, e.g., with general structure $u_k = k(x_k)+\bar{u}_k$, with a corresponding differential form as in \eqref{equ:pre dif fed}, i.e., $\frac{\partial k}{\partial x} = K$. Then, substituting \eqref{equ:pre dif fed} into \eqref{eq:dsys_dist_nom} yields, the closed-loop differential system for \eqref{eq:sys_dist_nom} as
    \begin{equation}\label{eq:diffcl_d}
        \delta_{x_{k+1}} = \left(A(x_k) + B_u K(x_k)\right)\delta_{x_k} + B_\nu\delta_{\nu_k}.
    \end{equation}
Substituting the closed-loop differential dynamics \eqref{eq:diffcl_d} and differential Lyapunov function,
\begin{equation}\label{equ:generalised distance}
    V(x_k,\delta_{x_k}) = \delta_{x_k}^\top M(x_k)\delta_{x_k},
\end{equation}
into the exponential discrete-time Lyapunov condition of 
\begin{equation}
    \label{eqe:Vdotcondition}
    V_{k+1} - (1-\beta)V_k < 0,
    \end{equation}
provides the contraction condition in \eqref{eq:dist ccon}. To see that satisfaction of this condition implies stabilisability, we proceed as follows. 
Integrating \eqref{eqe:Vdotcondition} with \eqref{equ:generalised distance} along the geodesic $\gamma(x_k, x_k^*)$ yields
\begin{equation}\label{inequ:distance bound2}
d^2(\gamma(x_{k+1},x^*_{k+1})) \leq (1-\beta) d^2(\gamma(x,x^*))
\end{equation}
using \eqref{equ:Riemannian distance and energy}, \eqref{equ:geodesic} and the property $d(\gamma)^2 \leq d(c)^2$. 
\end{proof}
\begin{remark}
Provided a transformation matrix $\Theta$ can be found (see \cite{lohmiller1998contraction}), such that $B_\nu\delta_{\nu} = \Theta \delta_x$ $\forall \delta_x \in \mathcal{X},\delta_\nu \in \mathcal{V}$, then \eqref{eq:dist ccon} can be expressed as
\begin{equation}
\begin{split}
    (A_{c} + \Theta)^\top M(x_{k+1})(A_{c} + \Theta) - (1-\beta)M(x_k)
      < 0,
\end{split}
\end{equation}
which can be transformed into a tractable synthesis condition using the techniques in \cite{wei2021control}. Moreover, this condition further simplifies in the case where the exogenous input is known.
\end{remark}
\begin{remark}
The term $\delta_v = \frac{\partial \nu}{\partial s}$  exists when the actual exogenous input $\nu$ (at $s=0$) varies away from the modeled input $\nu^*$ (at $s=1$). If the modeled trajectory is updated at each sampling instant, $k$, then $\nu^*_k =\nu_k$ and $\delta_{\nu_k} = 0$.
\end{remark}
\begin{corollary}\label{cor:cccontroller}
If a pair of matrix functions $(W(x_k),L(x_k))$ can be found satisfying the tractable condition
\begin{equation}\label{eq:tractable_con}
 \begin{bmatrix}
            W_+ & AW+B_u L \\
           (AW+B_u L)^\top  & (1-\beta)W
        \end{bmatrix} > 0,
\end{equation}
where $W = M^{-1}$, $L=KW$, $W_+=W(x_{k+1})$, $M$ is uniformly bounded as $\underline{m}I \leq M(x) \leq \overline{m}I$ and $\beta \in (0,1]$, then $M$ is a discrete-time-control contraction metric and $K$ is a contracting differential feedback gain. Moreover, a contracting tracking controller is given by \eqref{equ:pre ctl int} which ensures exponential convergence of $x$ to $x^*$ for \eqref{eq:sys_dist_nom}.
\end{corollary}
\begin{proof}
If $\nu_k$ is measured at each sampling instant, $k$, $\nu^*_k =\nu_k$ and $\delta_{\nu_k} = 0$. Hence, under exact disturbance measurement, the contraction condition in \eqref{eq:dist ccon} reduces to
\begin{equation}\label{equ:ccon_2} 
        (1-\beta)M - (A+B_u K)^\top M_+ (A+B_u K) > 0.
    \end{equation}
Defining $W(x_k) := M^{-1}(x_k)$, $W(x_{k+1}) := M^{-1}(x_{k+1})$, $L(x_k) := K(x_k)W(x_k)$, applying Schur's complement to \eqref{equ:ccon_2}, left/right multiplying by an invertible positive definite matrix, $\diag\{I,W\}$ (and its transpose) yields the equivalent tractable state-dependent linear matrix inequality \eqref{eq:tractable_con}. 

Since $M$ is uniformly bounded, from \eqref{equ:Riemannian distance and energy} and \eqref{equ:geodesic} we have
    \begin{equation}\label{inequ:a1 a2 bound}
        \underline{m}|x_k-x^*_k|^2 \leq d^2(\gamma_k),\ \ 
        d^2(\gamma_0) \leq \overline{m}|x_0-x^*_0|^2.
    \end{equation}
Combining \eqref{inequ:distance bound2} and \eqref{inequ:a1 a2 bound}, and taking the square root gives
    \begin{equation}
        |x_k-x^*_k| \leq (1-\beta)^{\frac{k}{2}} \sqrt{\frac{\alpha_2}{\alpha_1}} |x_0 - x^*_0|,
    \end{equation}
and hence exponential incremental stability for feasible trajectories of \eqref{eq:sys_dist_nom}. Finally, the contracting tracking controller \eqref{equ:pre ctl int} is obtained as in Section \ref{sec:pre}, by integrating the contracting differential feedback law \eqref{equ:pre dif fed} along the geodesic, $\gamma(x,x^*)$.
\end{proof}
\begin{remark}
To see that the state-dependent linear matrix inequality in \eqref{eq:tractable_con} is computationally tractable, and for a systematic synthesis method we refer to \cite{wei2021control}.
\end{remark}

\subsection{Analysis under Bounded Disturbance Prediction Errors}
If disturbance prediction is not exact, the $\mathcal{L}_2$ gain of the closed-loop system from the disturbance prediction error to the state error is attenuated.
\begin{corollary}\label{cor:tracking offset}
The nonlinear system with exogenous input \eqref{eq:sys_dist_nom} can be driven exponentially by a contraction-based feedback controller to a ball about the desired reference trajectory, i.e.,
\begin{equation}
        \label{eq:bball}
        d(\gamma(x_{k+1}, x_{k+1}^*)) \leq (1-\beta)^\frac{1}{2}d(\gamma(x_k, x_k^*)) + \sqrt{\overline{m}}G_k \|\tilde{u}_k\|.
    \end{equation} 
Moreover, this ball shrinks to zero for the case when the exogenous input is exactly measured and modelled correctly.
\end{corollary}
\begin{proof}
Suppose the exogenous input dynamics can be directly incorporated into a stabilising feedback controller (e.g., \eqref{equ:pre ctl int}) through updates on the feedforward component, i.e., given the targets $(x^*_{k+1},x^*_k)$ solve $x^*_{k+1} = f(x^*_k) + g_u(x_k^*)u^* +  g^*_\nu(x_k)\nu^*$ for $u^*$. In this exact modelling setting, $\nu^*=\nu$, implies $\delta_{v} = 0$, and hence from the result in Proposition \ref{prop:ccon_no synth}, i.e., from \eqref{inequ:distance bound2}, we have
\begin{equation}\label{eq:shrinking_Rdist}
    d(\gamma(x_{k+1}, x_{k+1}^*)) \leq (1-\beta)^\frac{1}{2}d(\gamma(x_k, x_k^*)). 
\end{equation}
From the definition of a metric function and \eqref{eq:shrinking_Rdist}, for \eqref{eq:sys_dist_nom} we have (see, e.g., \cite{wei2021discrete})
\begin{equation}\label{equ:disturbance distance}
\begin{aligned}
    &d(\gamma(x_{k+1}, x_{k+1}^*)) = d(\gamma(\check{x}_{k+1} + g_u(x_k)\tilde{u}_k, x_{k+1}^*)) \\
    & \leq d(\gamma(\check{x}_{k+1}, x_{k+1}^*)) + d(\gamma(\check{x}_{k+1} + g_u(x_k)\tilde{u}_k, \check{x}_{k+1})),
\end{aligned}
\end{equation}
where the dynamics $\check{x}_{k+1} := f(x_k) + g_u(x_k)\check{u}_k + g^*_\nu(x_k)\nu^*_k$ have $(\check{x}_{k+1},x^*_{k},\check{u}^*_{k})$ as a feasible target solution, $\tilde{u}_k = \check{u}^*(x^*,\nu^*) - u^*(x^*,\nu)$ denotes the deviation due to exogenous input modelling in control solutions with $(x^*_{k+1},x^*_{k},u^*_{k})$ as a feasible target solution to \eqref{eq:sys_dist_nom}. Since the metric, $M_k$, is bounded by definition, then, 
\begin{equation}\label{eq:max_deviations}
   d(\gamma(\check{x}_{k+1} + g(x_k)\tilde{u}_k, \check{x}_{k+1})) \leq \sqrt{\overline{m}}G_k \|\tilde{u}_k\|, 
\end{equation}
where $G_k = \max_{x_k} \| g_u(x_k) \|$. From \eqref{eq:shrinking_Rdist}--\eqref{eq:max_deviations} we have \eqref{eq:bball}.
\end{proof}

In the following, we adapt differential dissipativity results \cite{wei2021control} to derive an $\mathcal{L}_2$ gain bound condition of the corresponding closed-loop system from the disturbance prediction error to the state error.
\begin{proposition}\label{prop:dd_cc}
For a sequence of $T+1$ steps disturbance predictions (not exact), the closed-loop system \eqref{eq:sys_dist_nom}, \eqref{eq:dsys_dist_nom}, \eqref{equ:pre dif fed} is exponentially stable with the incremental truncated $\mathcal{L}_2$ gain from $\nu$ to $x$ bounded by $\alpha$  over the finite interval $k \in [0,T]$, for any feasible  reference, $x_k^*$, and actual  disturbance, $\nu_k^*$,
\begin{equation}\label{eq:bounded dist response}
    \sum_{k=0}^T ||x_k - x_k^* ||_2 \leq \alpha \sum_{k=0}^T||\nu_k - \nu_k^* ||_2,
\end{equation} 
provided a pair, $(M,K)$, can be found satisfying 
\begin{equation}\label{equ:diff diss sub}
\begin{split}
    \delta_{x_k}^\top A_{c}^\top M_+ A_{c}\delta_{x_k} - (1-\beta)\delta_{x_k}^\top M \delta_{x_k}
    \\+ \delta_{x_{k}}^\top A_{c}^\top M_+ B_{\nu}\delta_{\nu_k} 
    + \delta^\top_{\nu_k} B^\top_{\nu} M_+ A_{c}\delta_{x_{k}}
    \\+ \delta^\top_{\nu_k} B^\top_{\nu} M_+ B_{\nu}\delta_{\nu_k} \leq \delta_{x_{k}}^\top Q \delta_{x_{k}} + 2 \delta_{\nu_{k}}^\top S \delta_{x_{k}} + \delta_{\nu_{k}}^\top R \delta_{\nu_{k}}.
\end{split}
\end{equation}

\end{proposition}
\begin{proof}
System~\eqref{eq:sys_dist_nom} with differential dynamics \eqref{eq:dsys_dist_nom} is exponentially differentially ($Q,S,R$)-dissipative if there exists a differential storage function $V(x,\delta_x)$ such that
\begin{equation}\label{eq:diff_diss}
    V_{k+1} - (1 - \beta) V_k\leq \begin{bmatrix}
\delta_x \\ \delta_\nu
\end{bmatrix}^\top
\begin{bmatrix} Q & S\\S^\top & R \end{bmatrix}
\begin{bmatrix}
\delta_x \\ \delta_\nu
\end{bmatrix}.
\end{equation}
Substituting the closed-loop differential dynamics \eqref{eq:diffcl_d} and generalised squared distance \eqref{equ:generalised distance} (as a differential storage function) into the exponential differential dissipativity condition of \eqref{eqe:Vdotcondition} provides the  exponential differential $(Q,S,R)$-dissipativity condition in \eqref{equ:diff diss sub}. The condition in \eqref{eq:bounded dist response} can then be concluded from \eqref{equ:diff diss sub} with $\alpha = ||\hat{Q}^{-\frac{1}{2}}||_2(||\hat{Q}^{-\frac{1}{2}} S||_2 + (R+S^\top \hat{Q}^{-1}S)^{\frac{1}{2}})$ using the results in \cite{wei2021contractionsyntehsis}.
\end{proof}
\begin{corollary}\label{cor:dd_synth}
If a pair of matrix functions $(W(x_k),L(x_k))$ can be found satisfying the tractable condition
\begin{equation}\label{equ:LMI_dd}
\begin{bmatrix} 
W_+ & A W + B_u L & B_{\nu} & 0\\
(A W + B_u L)^\top & (1 - \beta)W & 0 & W \\ 
B_{\nu}^\top & 0 & \alpha^2 I & 0\\
0 & W & 0 & I 
\end{bmatrix}  \geq 0,
\end{equation}
then the incremental truncated $\mathcal{L}_2$ gain from the disturbances to states is equal to $\alpha$, where $W = M^{-1}$, $L=KW$, $W_+=W(x_{k+1})$, and $\beta \in (0,1]$.
\end{corollary}
\begin{proof}
Choosing $Q=-I$, $S=0$ and $R =  \alpha^2 I$, \eqref{equ:diff diss sub} can be written as (with $A_c = A+B_u K$)
\begin{equation}
\begin{bmatrix} I + (1 - \beta)M & \!\!\!0\\0 & \!\!\!\alpha^2 I \end{bmatrix} -
[A_c,B_{\nu}]^\top M_+ [A_c,B_{\nu}]  \geq 0.
\end{equation}
By Schur's complement, defining $W(x_k) := M^{-1}(x_k)$, $W(x_{k+1}) := M^{-1}(x_{k+1})$, $L(x_k) := K(x_k)W(x_k)$, left/right multiplying by $\diag\{ I , W, I\}$ (and its transpose), and taking Schur's complement again,  gives \eqref{equ:LMI_dd}.
\end{proof}
\begin{remark}
To see that the state-dependent linear matrix inequality in \eqref{equ:LMI_dd} is computationally tractable, and for a systematic synthesis method we refer to \cite{wei2021contractionsyntehsis}.
\end{remark}
Solving for the pair $(M,K)$ in \eqref{equ:diff diss sub}, yields an incremental truncated $\mathcal{L}_2$ gain equal to $\alpha$, providing an additional design variable (implicit to the contraction-based controller of Corollary \ref{cor:cccontroller}) with respect to disturbance measurement error rejection. The result from the above analysis is that in the presence of disturbance measurement error, a contraction-based controller will still drive the system exponentially to a neighbourhood about the desired trajectory. Moreover, a bounded tracking response (incremental truncated $\mathcal{L}_2$ gain) can be ensured with respect to bounded disturbance measurement errors and imposed during metric-controller synthesis.

\subsection{Contraction-based Stability Constraint}
Optimal solutions to the MPC problem \eqref{eq:EMPCini} (without the constraint $r \leq 0$) can be associated with state values which deviate from target operating points. As a consequence, an MPC requires a contraction based constraint or equivalent to ensure reference tracking, and is constructed as follows.

\begin{corollary}\label{cor:cc_predH}
For \eqref{eq:sys_dist_nom}, the contraction constraint 
\begin{equation}\label{eq:cc_predH}
\begin{split}
    &r(\hat{x}_i,x^*_i,\hat{\nu}_i,\hat{u}_i,\beta):= d\left(\gamma(\hat{x}_{i+1},x^*_{i+1})\right) \\& \qquad - \left(1-\beta) \right)^{(i+1)/2} d\left(\gamma(\hat{x}_0,x^*_0)\right) \leq 0, 
\end{split}
\end{equation}
where future states $\hat{x}_{i+1}(\hat{x}_{i},\hat{\nu}_{i},\hat{u}_{i})$ are generated iteratively, defines the set of contracting tracking controllers which ensure the system state, $\hat{x}_i$, is exponentially convergent to the feasible trajectory, $x^*_i$, for each prediction step $i$.
\end{corollary}
\begin{proof}
Straightforward from the proof of Proposition \ref{prop:ccon_no synth}, Corollary \ref{cor:cccontroller} and Corollary \ref{cor:tracking offset}.
\end{proof}
\begin{remark}
The contraction constraint in \eqref{eq:cc_predH}, has a reference-independent structure and is a nonlinear function affine in the control action, $\hat{u}$. To see this, consider that at each $i$-th prediction step, the Riemannian geodesic distance, $d(\gamma(x_i,x^*_i)$, depends on the path $\gamma(\hat{x}_i,x_i^*)$, which is computed iteratively using the model $\hat{x}_{i+1} = f(\hat{x}_i) + B_u\hat{u}_i + B_\nu \hat{\nu}_i$, the reference trajectory $x_i^*$ and \eqref{equ:geodesic}, which depend on the current state measurement, $x_0$, current disturbance measurement, $\nu^*_0$, and predicted disturbance trajectory $\boldsymbol{\hat{\nu}} = (\nu^*_0,\hat{\nu}_1,\cdots,\hat{\nu}_{N-1})$, yielding the predicted control sequence $\boldsymbol{\hat{u}} = (\hat{u}_0,\hat{u}_1,\cdots,\hat{u}_{N_-1})$.
\end{remark}

The information required for the computation of $d(\gamma)$ in \eqref{eq:cc_predH} is available from the current state and disturbance measurements with knowledge of the system model and reference dynamics. In addition, this condition define a set (or the conditional property) of controllers which guarantee the system is contracting (under Proposition \ref{prop:ccon_no synth}), whereby one particular feasible control solution is given by the state-feedback control structure proposed Corollary \ref{cor:cccontroller}. For the case where the disturbance measurements are not exact, additional restrictions can be made during metric synthesis, e.g., those proposed in Corollary \ref{cor:dd_synth}, to ensure boundedness of the impact of the disturbance measurement error on the reference tracking error, as per Proposition \ref{prop:dd_cc}.

Using a feedback controller, e.g., \eqref{equ:pre ctl int}, the concept of feedforward compensation control can be leveraged, such that the desired reference input $u^*$ can be updated to account for exogenous input signals, providing a limited avenue for compensation, with a heavy reliance of measurement accuracy and potentially large rates of input actuation (i.e., ``aggressive control''). For example, in the simplest case with $B_u = B_v = I$, and the disturbance is known exactly, i.e., $\nu^* = \nu$, assign $u^* = u^*-\nu$.
This however, yields a sub-optimal solution, since the future trajectories of disturbances are not considered and motivates direct consideration of the known exogenous dynamics, including predictions, for analysis and control. As can be seen in \eqref{eq:cc_predH}, there is no need for solving \eqref{eq:sys_dist_nom} for the reference input $u^*$ when imposed on the MPC \eqref{eq:EMPCini}, reducing measurement accuracy reliance and large actuation rates by avoiding feedforward compensation and considering the full (predicted) disturbance trajectory. 

\section{Example} \label{sec:exa}
Consider the following discrete-time Lotka-Volterra model
\begin{equation}\label{sim:sys}
\begin{split}
x_{1,k+1} &= (1+\tau \alpha) x_{1,k} + \tau \beta x_{1,k} - \tau \alpha  x_{1,k} x_{2,k} + u_k + \nu_k\\
x_{2,k+1} &= (1 - \tau \alpha) x_{2,k} + \tau \beta x_{2,k} +\tau \alpha x_{1,k} x_{2,k},
\end{split}
\end{equation}
where $x_1 \in [0.1,2]$ is the predator (product), $x_2 \in [0.1,2]$ is the prey (energy/food source), $u \in [-1,1]$ is the human interaction (via product manipulation) and $\nu_k = 0.1 \sin (0.1 k) + 0.1w_k$ is an environmental disturbance (uncontrolled, exactly measured) where $w_k$ is normally distributed Gaussian noise. Normalised growth and decay parameters are modeled with $(\alpha,\beta) = (1,0.001)$, and the sampling rate is $\tau = 0.1$. The goal is a stable ecosystem with minimal human interaction. 

Firstly, a discrete-time contraction metric, $M$, was synthesised with rate $\beta = 0.1$ using \eqref{eq:tractable_con} (see \cite{wei2021contractionsyntehsis}). The MPC in \eqref{eq:EMPCini} was constructed with prediction horizon, $N=5$, and squared norm cost function, $\ell = ||u_k||^2$. The target operating condition is a stable (periodic) orbit generated using \eqref{sim:sys} with the state initial condition $x_0 = (0.99,0.99)$, and $u = \nu = 0$. The predicted disturbance is incorrectly modeled with $w=0$. 

\begin{figure}\label{fig:sim}
    \begin{center}
        \includegraphics[width=\linewidth]{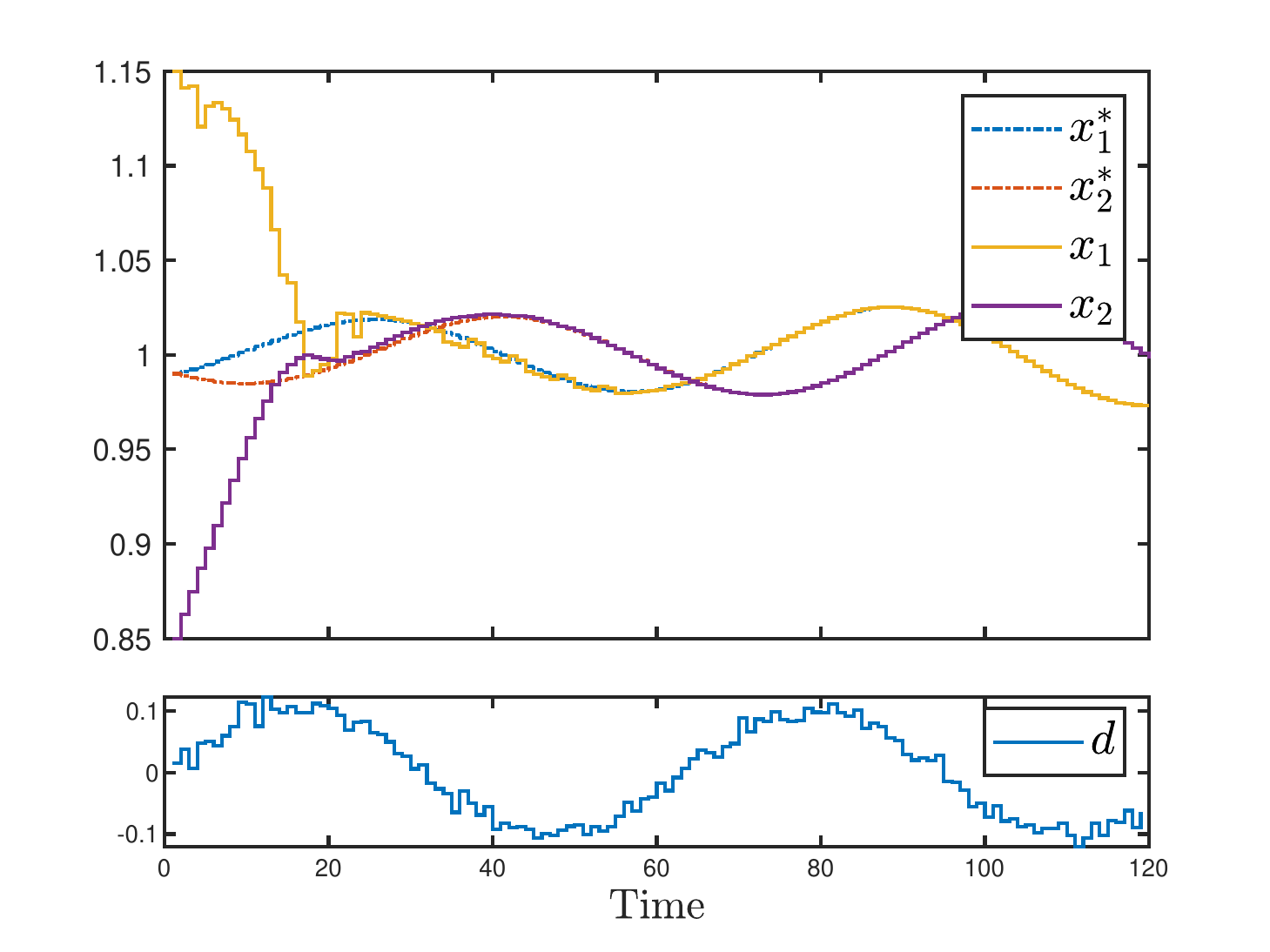}
        \caption{Simulated Response.}
        \label{fig:sim}
    \end{center}
\end{figure}

Fig. \ref{fig:sim} shows the simulated states tracking their time-varying references (stable ecosystem oscillations) despite an incorrectly predicted (accurately measured) disturbance.
\section{Conclusion} \label{sec:conclusion}
A contraction constrained MPC approach was developed for discrete-time nonlinear processes subject to predicted disturbances. Through the contraction theory framework, stability conditions were derived to ensure convergence of the resulting closed-loop control system. Computationally tractable equivalent conditions were also derived, such that joint synthesis of a discrete-time control contraction metric and stabilising feedback controller could be completed with desired closed-loop properties. These conditions were also generalised to characterise a set of stabilising controllers and imposed as stability constraints on an MPC. The MPC would then search for a cost minimising controller amongst those stabilising controllers, whilst utilising both forecasts and measurements. The result was an MPC capable of ensuring convergency of nonlinear processes subject to measured disturbances, as demonstrated through a numerical case study.

\bibliographystyle{IEEEtran}
\bibliography{AdCONIP22_main.bib}

\end{document}